\documentclass[conference]{IEEEtran}
\IEEEoverridecommandlockouts
\usepackage{cite}
\usepackage{amsmath,amssymb,amsfonts}
\usepackage{algorithmic}
\usepackage{graphicx}
\usepackage{textcomp}
\usepackage{xcolor}
\usepackage{url}
\def\BibTeX{{\rm B\kern-.05em{\sc i\kern-.025em b}\kern-.08em
    T\kern-.1667em\lower.7ex\hbox{E}\kern-.125emX}}
\DeclareRobustCommand*{\IEEEauthorrefmark}[1]{%
    \raisebox{0pt}[0pt][0pt]{\textsuperscript{\footnotesize\ensuremath{#1}}}}
\begin{document}

\title{HSD: A hierarchical singing annotation dataset}

\author{
\IEEEauthorblockN{
Xiao Fu\IEEEauthorrefmark{1},
Xin Yuan\IEEEauthorrefmark{1},
Jinglu Hu\IEEEauthorrefmark{1}}
\IEEEauthorblockA{\textit{\IEEEauthorrefmark{1}Graduate School of Information Production
and Systems}\\
\textit{Waseda University}\\
\textit{Kitakyushu, Fukuoka, Japan}}
\IEEEauthorblockA{\textit{hirabara@toki.waseda.jp, sherryyuan@ruri.waseda.jp, jinglu@waseda.jp}}}


\maketitle

\begin{abstract}
Commonly music has an obvious hierarchical structure, especially for the singing parts which usually act as the main melody in pop songs.
However, most of the current singing annotation datasets only record symbolic information of music notes, ignoring the structure of music.
In this paper, we propose a hierarchical singing annotation dataset that consists of 68 pop songs from Youtube.   
This dataset records the onset/offset time, pitch, duration, and lyric of each musical note in an enhanced LyRiCs format to present the hierarchical structure of music.
We annotate each song in a two-stage process: first, create initial labels with the corresponding musical notation and lyrics file; second, manually calibrate these labels referring to the raw audio.
We mainly validate the labeling accuracy of the proposed dataset by comparing it with an automatic singing transcription (AST) dataset.
The result indicates that the proposed dataset reaches the labeling accuracy of AST datasets.
\end{abstract}


\section{Introduction}
\begin{figure}[!t]%
    \centering
    \includegraphics[width=3.4in]{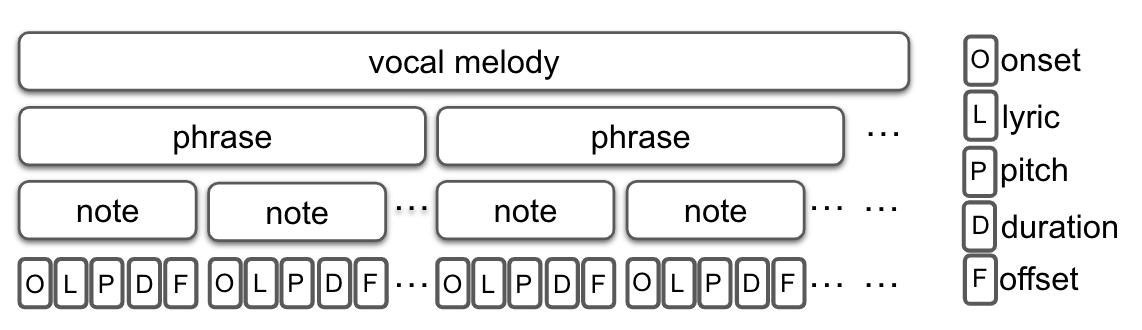}
    \caption{The hierarchical structure of vocal music.}
    \label{fig1}
\end{figure}

Vocals in pop songs usually have a hierarchical structure as described in Figure \ref{fig1}.
A piece of vocal music usually consists of several musical phrases and each phrase contains several musical notes.
Meanwhile, a single musical note presents five basic properties: onset, pitch, duration, lyric, and offset.
Recent years have been increased interest in singing-related tasks due to their potential value in both scientific and commercial applications.
Subsequently, various singing annotation datasets have been proposed to assist these researches.
According to the research motivations, the contents of these singing datasets are also different.
For instance, AST \cite{nishikimi2019automatic} mainly requires the data of pitch, onset, and offset time for each musical note, symbolic music generation (SMG) \cite{MCTS} needs pitch and duration, and automatic lyric transcription (ALT) \cite{alt} calls for lyrics and the corresponding timestamps. 
However, there are few singing annotation datasets that fully record the hierarchical information.

In this work, we present a singing annotation dataset (namely HSD), which not only provides the information of onset/offset, pitch, duration, and lyric for each musical note but also records these data in a hierarchical structure.
We initialize the vocal labels via a method (named N\&L) using the corresponding musical notation and LyRiCs (LRC) file and then execute a manual calibration process to produce the final labels.
We record the labeled data in an enhanced-LRC format to present the hierarchical structure of music.
We validate the labeling accuracy by comparing it with the MIRST-500 dataset \cite{MIRST500} which is proposed for AST task.

\section{Related Work}
In the last few decades, several vocal-music-related datasets have been proposed for various MIR tasks.
The Jamendo lyrics database \cite{jamendolyrics} was designed for lyrics alignment, transcription, and evaluation.
The iKala dataset \cite{chan2015vocal} and MIR-1K dataset \cite{mir1k} were offered for singing voice separation.
The ISMIR 2014 dataset \cite{ISMIR2014}, TONAS dataset \cite{tonas}, and MIR-ST500 dataset \cite{MIRST500} were provided for AST. 
The VocalSet dataset \cite{wilkins2018vocalset} contained monophonic recorded audio of professional singers and was designed for singer identification, vocal technique identification, singing generation, and other related applications.
The RWC music database \cite{goto2002rwc}, which labeled both the vocals and the instrumental accompaniment in real-world songs, was proposed for common use and for the purposes of musical information processing research.

We compare the contents of HSD with several similar datasets as shown in Table \ref{table:1}.
The MIR-ST500 dataset \cite{MIRST500} labels the pitch, onset/offset time of each musical note.
The Jamendo lyrics database \cite{jamendolyrics} labels each lyric and its onset/offset time, besides the timestamp for each phrase.
The RWC music database \cite{goto2002rwc} provides comprehensive annotations for each musical note whereas the lyrics are recorded at the phrase level.
The proposed dataset not only labels the lyric, pitch, duration, and onset/offset time but also records these data in a hierarchical format for potential exploitation in the future.

\begin{table*}[!t]
\footnotesize
\renewcommand{\arraystretch}{1.3}
\caption{Content Comparison of HSD with Similar Datasets}
\label{table:1}
\centering.
\begin{tabular}{|c|c|c|c|c|c|c|c|}
\hline
Dataset & Lyric & Pitch & Duration & Onset/Offset & Phrase-Timestamp & Hierarchical structure\\

\hline
MIR-ST500 & \(\times\) & $\checkmark$ & \(\times\) & $\checkmark$ & \(\times\) & \(\times\) \\
\hline
Jamendo lyrics & note-level lyrics & \(\times\) & \(\times\) & $\checkmark$ & $\checkmark$ & \(\times\) \\
\hline
RWC & phrase-level lyrics & $\checkmark$ & $\checkmark$ & $\checkmark$ & $\checkmark$ & \(\times\) \\
\hline
HSD (Ours) & note-level lyrics & $\checkmark$ & $\checkmark$ & $\checkmark$ & $\checkmark$ & $\checkmark$ \\

\hline
\end{tabular}
\end{table*}

\section{Labeling Method}
The labeling process is executed in two-step: initialization and calibration.
We use an N\&L (i.e., notation and LRC) method to create the initial labels.
For given raw audio, we first collect the corresponding musical notation and LRC file from the web (the music notation offers pitch and duration for each music and LRC file provides phrase-level lyrics and timestamps).
Second, we segment the raw audio into music phrases according to the phrase onset time tags $t_{1:n}$ from the LRC file, where $n$ represents the phrase number.
At last, we locate each musical note by computing its percentage in the phrase according to the duration $d_{1:m}$ from the musical notation, where $m$ represents the number of notes in the sentence.
Then the onset time $O$ and offset time $F$ of each musical note are calculated as follows.
\begin{equation}
O_{\mu\nu} = t_\nu + (t_{\nu+1} - t_{\nu}) \frac{\sum\limits_{i=1}^{\mu-1}d_i}{\sum\limits_{i=1}^{m}d_i} \label{eq_on}
\end{equation}
\begin{equation}
F_{\mu\nu} = O_{\mu\nu} + (t_{\nu+1} - t_{\nu}) \frac{d_{\mu}}{\sum\limits_{i=1}^{m}d_i} \label{eq_off}
\end{equation}
Here, $\mu$ and $\nu$ are the indices of the note and sentence, respectively.
This initialization method requires the annotators to have some basic knowledge of music to read the musical notation and input each musical symbol into the initial program.
We have not utilized a baseline model to automatically create the initial labels just as Molina \cite{ISMIR2014} and Wang \cite{MIRST500} did for AST datasets because training such a model to predict all the needed data is difficult.

After the initialization, a manual calibration process is executed to correct the coarsely initialized labels to accurate labels.
The initialized labels are processed to MIDI format first and then put into a parallel track with the raw audio track.
The annotators play these two tracks simultaneously to compare the vocals with the annotations.
Since the pitch and duration obtained from musical notation are almost correct, the annotators are mainly needed to fix the data of the onset/offset time.
We do not execute the calibration note-by-note, but instead, we mainly calibrate the onset time tags $t$ of each phrase because the onset/offset time of each musical note is determined by the phrase timestamp $t$ (Eq. \ref{eq_on}-\ref{eq_off}).
After two annotators (one for labeling and another for validation) consider the labels are already accurate, the annotations of a song are exported to an enhanced LRC format as the final output.

Compared with the methods by Molina \cite{ISMIR2014} and Wang \cite{MIRST500} which use a baseline model for initialization and human examination for calibration, our method shows some advantages and disadvantages.
First, manually creating the initial labels is more expensive and time-consuming than automatic transcription but can obtain more data such as duration, lyrics, and the hierarchical structure.
Second, the pitch obtained from the music notation is almost correct thus the annotators can focus on fixing the onset/offset.
But in some rare cases, the raw audio includes a short piece of improvised humming that is not recorded in the musical notation and LRC file.
Therefore this kind of vocals can not be initialized and labeled by our method.
Third, in the calibration process, we fix the onset/offset of each musical note via adjusting the phrase timestamp, which is easier than note-by-note correction.

\section{Data Record}
The proposed dataset is available at \url{https://github.com/hirabarahyt/HSD-Dataset}.
There are 68 annotation files in enhanced LRC formats in the dataset; 62 are transcribed from Chinese songs, whereas the rest are transcribed from Japanese songs.
The Youtube link for each song is offered.
The enhanced LRC format is based on the ordinary LRC format with an onset time tag ``$\langle$mm:ss.xx$\rangle$'' and an offset time tag ``$\{$mm:ss.xx$\}$'' added for each lyric, where ``mm'' is minutes, ``ss'' is seconds, and ``xx'' is hundredths of a second.
We included the pitch and duration alongside each lyric in the format ``l p d'' after the onset time tag and before the offset time tag, where ``l'' is the lyric, ``p'' is the pitch, and ``d'' is the duration.
We used simplified Chinese characters to present lyrics for Chinese songs and Katakana for Japanese songs.
The duration was transformed to floating point units, for example, a quarter note typically represents one beat, and a sixteenth note is one-quarter of a quarter note, so a sixteenth note duration is represented as 0.25.
The phrase time tag is stated at the beginning of each phrase in the format ``[mm:ss.xx].''
Each line in the enhanced LRC file can thus be described as follows:
[mm:ss.xx]$\langle$mm:ss.xx$\rangle$l p d$\{$mm:ss.xx$\}$\ldots $\langle$mm:ss.xx$\rangle$l p d$\{$mm:ss.xx$\}$.

We also provide the fixed notations, LRC files, and the source codes used to create the annotations in the same repository.
Users can reuse this framework to produce their own data if they have musical notations and LRC files.
Meanwhile, we provide two ways to create the MIDI files used for the calibration.
The first is based on the collected music notations and LRC files.
In this case, the annotator can just correct the phrase timestamps in LRC files for the calibration.
The second is based on the enhanced LRC files created by the music notations and LRC files.
The annotator can perform a note-by-note calibration process by correcting the onset/offset of each musical note in the enhanced LRC files.

\section{Validation}
We mainly validate the proposed dataset by comparing the COnP with the MIRST-500 \cite{MIRST500} to confirm the labeling accuracy.
However, intuitively quantifying the quality of singing annotations is difficult.
Therefore we employ a contrastive method in the evaluation process as follows.
For a given raw audio $r$ and its annotation $a$, we first shift the onset/offset time of all the musical notes $t$ milliseconds earlier and later to $a_{-}$ and $a_{+}$.
Then we compare the compatibility of $r$ with $a$, $a_{-}$, and $a_{+}$ to find the most suitable annotation.
If $a$ is the best for $r$, we will consider the annotation $a$ accurate; otherwise $a$ will be considered inaccurate.
Since evaluating the whole dataset is time-consuming, we randomly select 5 songs from HSD and MIRST-500 respectively for the validation.
The results based on $t=50ms$ and $t=100ms$ are shown in Table \ref{table:result}.
From the results we can observe that our dataset reaches the same labeling accuracy as MIRST-500.
\begin{table}[h]
\footnotesize
\renewcommand{\arraystretch}{1.3}
\caption{Comparison Result between MIRST-500 and HSD (\%)}
\label{table:result}
\centering.
\begin{tabular}{|c|c|c|}
\hline
Dataset & COnP (50ms) & COnP (100ms) \\
\hline
MIRST-500 & 96.49 & 98.42 \\ 
\hline
HSD (ours) & 97.07 & 98.53 \\
\hline
\end{tabular}
\end{table}

\section{Conclusion}
In this paper, we propose a singing annotation dataset that records the pitch, duration, lyric, onset, and offset of each musical note in a hierarchical format.
Through a contrastive validation method, we found our dataset achieves the labeling accuracy of the MIRST-500 dataset \cite{MIRST500} which is designed for the AST task.
Meanwhile, we provide the codes for creating this dataset so that users can make their own dataset if they have musical notations and the corresponding LRC files.
In the future, we will explore how to use this kind of hierarchical information for MIR tasks.

\bibliographystyle{IEEEtran}
\bibliography{HSD}

\end{document}